\documentclass[hidelinks, 10pt]{article}

\usepackage[utf8]{inputenc}
\usepackage{lmodern}         
\usepackage{geometry}
\usepackage{setspace}

\usepackage{graphicx}
\usepackage{subcaption}
\graphicspath{{img/}}

\usepackage{float}
\usepackage{longtable}
\usepackage{array}
\usepackage{rotating}

\usepackage{multirow}
\usepackage{makecell}
\usepackage{ragged2e}
\usepackage{booktabs}
\usepackage{listings} 
\usepackage{amsmath, amssymb, amsfonts}
\usepackage{fancyvrb}
\usepackage{verbatim}
\usepackage{fvextra}
\usepackage{enumitem}

\usepackage{titlesec}
\usepackage{tocloft}

\usepackage{caption}
\usepackage{hyperref}

\usepackage{etoolbox}
\AtBeginEnvironment{tabular}{\fontsize{10pt}{12pt}\selectfont}
\AtBeginEnvironment{longtable}{\fontsize{10pt}{12pt}\selectfont}
\AtBeginEnvironment{table}{\fontsize{10pt}{12pt}\selectfont}

\usepackage{ragged2e}

\titleformat{\section}[block]
  {\centering\normalfont\fontsize{14pt}{16pt}\bfseries\selectfont}
  {\thesection.}{0.5em}{}  
\titlespacing{\section}{0pt}{5pt}{0pt} 

\titleformat{\subsection}[block]
  {\normalfont\fontsize{12pt}{12pt}\bfseries\selectfont}
  {\thesubsection.}{0.5em}{}  
\titlespacing{\subsection}{0pt}{5pt}{0pt}

\titleformat{\subsubsection}[block]
  {\normalfont\fontsize{11pt}{11pt}\bfseries\selectfont}
  {\thesubsubsection.}{0.5em}{}  
\titlespacing{\subsubsection}{0pt}{5pt}{0pt}

\titleformat{\paragraph}[block]
  {\normalfont\fontsize{10pt}{10pt}\bfseries\selectfont}
  {\theparagraph.}{0.5em}{}
\titlespacing{\paragraph}{0pt}{5pt}{0pt}

\setlist[enumerate]{itemsep=5pt, topsep=5pt, partopsep=0pt, parsep=0pt, labelsep=0.5em, left=0pt}

\usepackage[
  backend=biber,
  style=numeric-comp,
  sorting=nyt,
  sortcites=true,
  doi=true,        
  url=true,
  isbn=false
]{biblatex}

\usepackage{glossaries}
\makeglossaries
\newacronym{arf}{ARF}{Acute Respiratory Failure}
\newacronym{ards}{ARDS}{Acute Respiratory Distress Syndrome}
\newacronym{nlp}{NLP}{Natural Language Processing}
\newacronym{ehr}{EHR}{electronic health record}
\newacronym{ehrs}{EHRs}{electronic health records}
\newacronym{eicu}{eICU-CRD}{eICU Collaborative Research Database}
\newacronym{n3c}{N3C}{National COVID Cohort Collaborative}
\newacronym{llms}{LLMs}{Large Language Models}
\newacronym{llm}{LLM}{Large Language Model}
\newacronym{pasc}{PASC}{Post-Acute Sequelae of SARS-CoV-2 infection}
\newacronym{sars}{SARS-CoV-2}{SARS-CoV-2}
\newacronym{imv}{IMV}{Invasive Mechanical Ventilation}
\newacronym{nippv}{NIPPV}{Noninvasive Positive Pressure Ventilation}
\newacronym{hfni}{HFNI}{High-Flow Nasal Insufflation}
\newacronym{niv}{NIV}{Noninvasive Ventilation}
\newacronym{icu}{ICU}{Intensive Care Unit}
\newacronym{gpt}{GPT}{generative pre-trained transformer}
\newacronym{gpu}{GPU}{graphics processing unit}
\newacronym{gpus}{GPUs}{graphics processing units}
\newacronym{api}{API}{application programming interface}
\newacronym{apis}{APIs}{application programming interfaces}
\newacronym{bert}{BERT}{Bidirectional Encoder Representations from Transformers}
\newacronym{rnn}{RNN}{recurrent neural network}
\newacronym{rag}{RAG}{retrieval-augmented generation}
\newacronym{icd}{ICD-10}{International Classification of Diseases (version 10)}
\newacronym{sql}{SQL}{Structured Query Language}
\newacronym{roc}{AUROC}{area under the receiver operating characteristic}
\newacronym{prc}{AUPRC}{area under the precision-recall curve}
\newacronym{cocomo}{COCOMO}{Constructive Cost Model}
\newacronym{phi}{PHI}{Protected Health Information}
\newacronym{slms}{SLMs}{statistical language models}
\newacronym{iqr}{IQR}{Interquartile range}
\newacronym{cot}{CoT}{Chain-of-Thought}
\newacronym{pheona}{PHEONA}{Evaluation of PHEnotyping for Observational Health Data}
\newacronym{omop}{OMOP-CDM}{Observational Medical Outcomes Partnership Common Data Model}
\newacronym{framework}{SHREC}{SHifting to language model-based REal-world Computational phenotyping}
\newacronym{amia}{AMIA}{American Medical Informatics Association}
\newacronym{bleu}{BLEU}{Bilingual Evaluation Understudy}
\newacronym{rouge}{ROUGE}{Recall-Oriented Understudy for Gisting Evaluation}
\newacronym{impact}{IMPACT}{Intelligent Machine for Patient Accrual and Classification Tasks}

\newglossaryentry{snomed}{
  name=SNOMED CT,
  description={Systematized Nomenclature of Medicine—Clinical Terms}
}

\begin{document}

\begin{center}
\textbf{\fontsize{14pt}{14pt}\selectfont SHREC: A Framework for Advancing Next-Generation Computational Phenotyping with Large Language Models}

\vspace{1.5ex}

\noindent
Sarah A. Pungitore, MS$^1$, Shashank Yadav, MS$^2$, Molly Douglas, MD$^3$, Jarrod Mosier, MD$^3$, Vignesh Subbian, PhD$^2$ \\[1ex]
$^1$ Program in Applied Mathematics, The University of Arizona, Tucson, AZ \\
$^2$ College of Engineering, The University of Arizona, Tucson, AZ \\
$^3$ College of Medicine - Tucson, Tucson, AZ
\end{center}
\stepcounter{section}

\makeatletter
\renewcommand{\@biblabel}[1]{\hfill #1.}
\makeatother

\begin{refsection}

\subsection{Abstract}
Computational phenotyping is a central informatics activity with resulting cohorts supporting a wide variety of applications. However, it is time-intensive because of manual data review and limited automation. Since \acrshort{llms} have demonstrated promising capabilities for text classification, comprehension, and generation, we posit they will perform well at repetitive manual review tasks traditionally performed by human experts. To support next-generation computational phenotyping, we developed SHREC, a framework for integrating \acrshort{llms} into end-to-end phenotyping pipelines. We applied and tested three lightweight \acrshort{llms} (Gemma$2$ $27$ billion, Mistral Small $24$ billion, and Phi-$4$ $14$ billion) to classify concepts and phenotype patients using phenotypes for \acrshort{arf} respiratory support therapies. All models performed well on concept classification, with the best (Mistral) achieving an \acrshort{roc} of $0.896$. For phenotyping, models demonstrated near-perfect specificity for all phenotypes with the top-performing model (Mistral) achieving an average \acrshort{roc} of $0.853$ for single-therapy phenotypes. In conclusion, lightweight \acrshort{llms} can assist researchers with resource-intensive phenotyping tasks. Several advantages of \acrshort{llms} included their ability to adapt to new tasks with prompt engineering alone and their ability to incorporate raw \acrshort{ehr} data. Future steps include determining optimal strategies for integrating biomedical data and understanding reasoning errors.

\subsection{Introduction}
Computational, or electronic phenotyping, is a central informatics activity focused on defining, extracting, and validating meaningful clinical representations of digital data from  \acrfull{ehrs} and other relevant information systems.\cite{hripcsak_next-generation_2013, callahan_characterizing_2023} It is particularly fundamental to observational studies,  large-scale pragmatic clinical trials, and healthcare quality improvement initiatives, where standardized, computable phenotypes allow for robust cohort discovery and monitoring of real-world outcomes.\cite{banda_advances_2018} Computable phenotypes have been developed for a wide variety of clinical outcomes and conditions, including acute conditions such as acute kidney injury,\cite{ozrazgat-baslanti_development_2024} \acrlong{ards},\cite{li_rule-based_2021} and acute brain dysfunction in pediatric sepsis,\cite{alcamo_validation_2022} and chronic conditions such as breast cancer,\cite{neely_design_2022} hypertension,\cite{mcdonough_optimizing_2020} and \acrfull{pasc}.\cite{pungitore_computable_2024} They have also supported a variety of downstream tasks, including recruitment for clinical trials, development of clinical decision support systems, and hospital quality reporting.\cite{banda_advances_2018, tekumalla_towards_2024, shang_making_2019}

\vspace{5pt}

The process of developing computable phenotypes typically includes identification and construction of relevant data elements for classification and then application of an algorithm to produce the cohort(s) of interest.\cite{carrell_general_2024} Traditionally, these processes involve multiple time and resource-intensive tasks requiring manual data review, such as mapping of data elements to controlled vocabularies.\cite{shang_making_2019} Despite increased adoption of controlled vocabularies in \acrshort{ehr} systems and improvements in \acrfull{nlp} and machine learning methods, computational phenotyping remains complicated and costly.\cite{tekumalla_towards_2024, shang_making_2019} As a result, many of the desiderata for phenotyping identified over a decade ago are still relevant today, indicating the need for substantial improvements to these methods.\cite{hripcsak_next-generation_2013, wen_impact_2023} To demonstrate these issues, we highlight challenges in development of computable phenotypes for \acrshort{pasc}.\cite{pungitore_computable_2024} Since the phenotype definition was based on symptom presence, manual expert review of $6,569$ concepts was first required to determine which were relevant to the $151$ symptoms of interest. A series of data transformations were then applied to assess symptom presence relative to \acrlong{sars} infection. Any new dataset, especially one not mapped to a controlled vocabulary, would require further manual review of concepts, rework of the algorithm, or both.

\vspace{5pt}

Given the existing opportunities with computational phenotyping and the minimal overall progress towards methodological improvements, it is natural to consider what will drive the next significant enhancement, or ``next-generation" of phenotyping methods. In particular, with advances in machine and artificial intelligence, we also reconsider how much of the computational phenotyping process requires direct human involvement. The idea of human-machine synergy, with each component enhancing the abilities of the other, is fundamental to the field of informatics.\cite{friedman_fundamental_2009} However, this synergy has yet to be achieved in computational phenotyping since humans still perform a majority of the phenotyping tasks, including ones where machines may excel. Therefore, we propose exploring the potential of \acrfull{llms} for this domain. As a relatively new addition to biomedical research, \acrshort{llms} introduce a novel set of text analysis, comprehension, and generation capabilities that allow them to analyze and generate text in ways that previously were either only possible by humans or by extensively trained, topic-specific \acrshort{nlp} models.\cite{raiaan_review_2024} Additionally, since \acrshort{llms} are widely available as pretrained foundation models, they can be adapted to new tasks through prompt engineering alone, which is a more accessible and portable method of model adaptation compared to model retraining or domain adaptation methods.\cite{liu_pre-train_2023} Furthermore, biomedical fine-tuned models underperformed on clinical tasks when compared to general-use \acrshort{llms}, indicating that model retraining (a costly and time-intensive process) isn't even preferred for \acrshort{llm} adaptation.\cite{dorfner_evaluating_2025} Thus, the capabilities and advantages of \acrshort{llms} satisfy many of the current deficits in computational phenotyping methods, suggesting their potential as foundational tools for next-generation phenotyping.

\vspace{5pt}

While some studies have applied \acrshort{llms} to various clinical phenotyping tasks, none have explored the capability of \acrshort{llms} to improve computational phenotyping specifically. The clinical phenotyping tasks studied include entity extraction and matching in clinical text,\cite{baddour_phenotypes_2024, zelin_rare_2024} query generation for patient extraction,\cite{yan_large_2024} evaluation of hospital quality measures,\cite{boussina_large_2024} and creation of phenotype definitions from standardized vocabulary codes.\cite{tekumalla_towards_2024} When used to develop queries for identifying patients with type-2 diabetes mellitus, dementia, and hypo-thyroidism, \acrshort{gpt}-4 produced queries that still required substantial oversight from human reviewers to generate accurate cohorts.\cite{yan_large_2024} Additionally, when \acrshort{llms} were used to generate computable phenotypes based on standardized vocabularies, \acrshort{gpt}-4 only achieved an average accuracy of approximately $50\%$ on both code matching and on string matching when compared to the original definition.\cite{tekumalla_towards_2024} However, SOLAR $10.7$B only slightly underperformed human categorizations for hospital quality measures and even provided a better response than human review in $4$ out of $10$ cases when responses between humans and \acrshort{llms} differed.\cite{boussina_large_2024} Additionally, GPT-$4$o demonstrated perfect accuracy for classification of antibiotics from raw \acrshort{ehr} data.\cite{matos_ehrmonize_2025} Therefore, while \acrshort{llms} struggle with query and algorithm generation, even lightweight models have demonstrated the ability to categorize relevant clinical concepts from \acrshort{ehr} data, further indicating potential application of \acrshort{llms} for development of computable phenotypes.

\vspace{5pt}

Considering the opportunities in computational phenotyping methods and the novel capabilities of \acrshort{llms}, we applied and evaluated \acrshort{llms} to support computable phenotype development. We previously developed \acrshort{pheona} (\acrlong{pheona}), a framework specifically for evaluating \acrshort{llms} for computational phenotyping tasks.\cite{pungitore_pheona_2025} In this study, we expanded upon these methods to construct a broader view of next-generation phenotyping. The objectives of this study were thus the following:

\begin{enumerate}
    \item Develop \acrshort{framework} (\acrlong{framework}), a companion framework to \acrshort{pheona} that outlines how to integrate \acrshort{llms} into computational phenotyping. 
    
    \item Apply and demonstrate \acrshort{framework} using previously developed computable phenotypes for \acrfull{arf} respiratory support therapies.

    \item Highlight future work and next steps to encourage progress towards next-generation phenotyping methods.
\end{enumerate}

\subsection{Materials and Methods}
We first outline the development of \acrshort{framework} along with its individual components (Figure \ref{figure_1} and Table \ref{tab:p2_table1}) and then we discuss how we used \acrshort{llms} to perform various tasks for a specific phenotyping use case.

\subsubsection{Development of SHREC}
\paragraph{Theoretical Foundation}
\noindent
To understand issues with computational phenotyping, we revisited the \textit{Fundamental Theorem of Informatics}, which states optimized human-machine interactions should drive informatics methods,\cite{friedman_fundamental_2009} and \textit{distributed cognition}, which describes how overall cognitive load is shared between internal and external agents.\cite{hazlehurst_distributed_2008, patel_cognitive_2021} In traditional phenotyping, humans generally could not delegate tasks to external agents without costly, time-consuming, or even impossible modifications.\cite{tekumalla_towards_2024, shang_making_2019, raiaan_review_2024}  Therefore, humans were responsible for both repetitive and complex tasks despite not being as inherently well-suited for repetitive work as machines. If \acrshort{llms} are indeed capable of performing repetitive tasks well, these tasks can be offloaded to external \acrshort{llm}-based agents while ensuring humans are only responsible for complex ones to reduce overall cognitive burden of researchers and improve efficiency of the phenotyping process.

\begin{figure}[ht]
  \centering
    \caption{An overview of \acrshort{framework} (\acrlong{framework}), including both an overview of the individual phenotyping tasks and a representation of the progress required to advance next-generation computational phenotyping.}\label{figure_1}
    \includegraphics[width=1.0\textwidth]{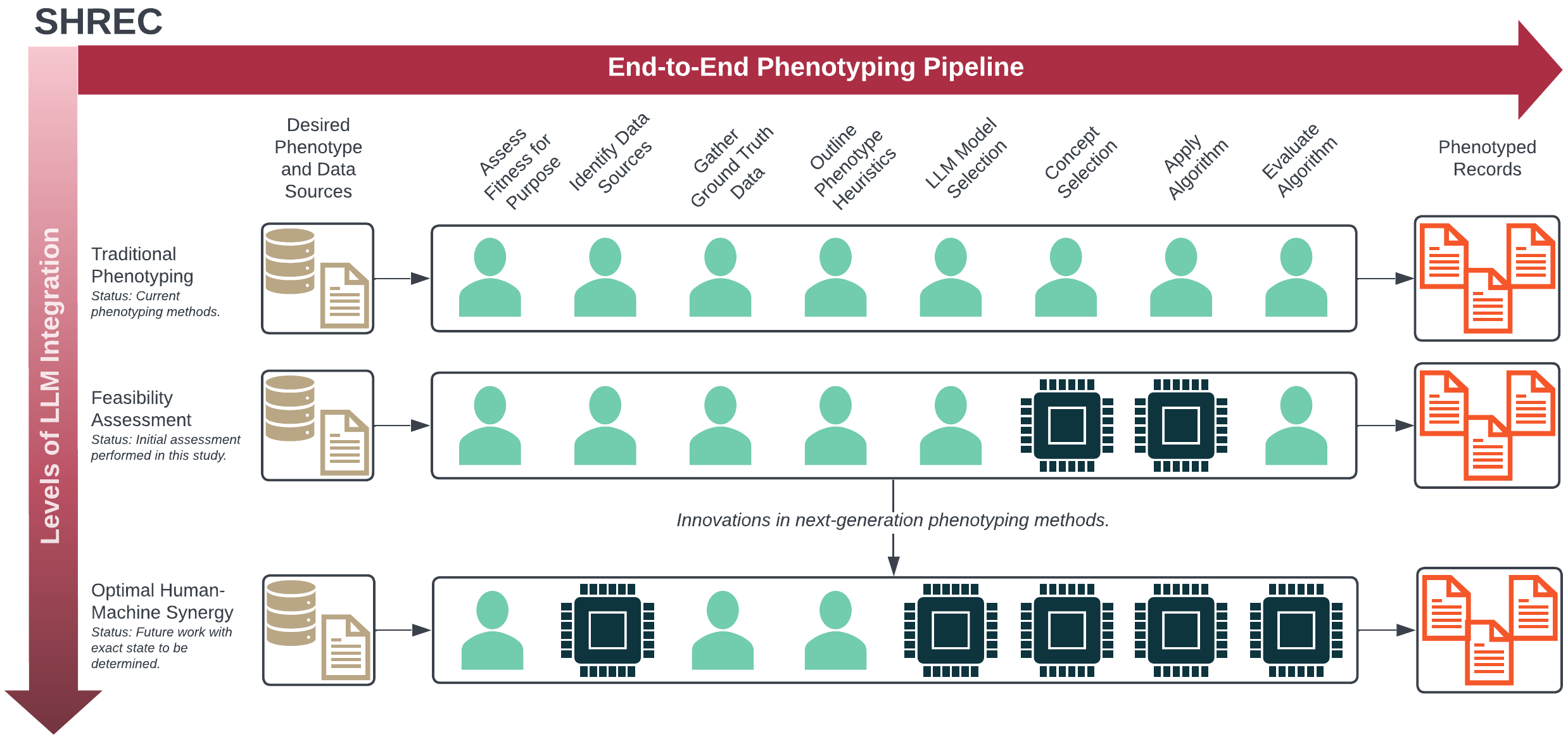}
 \label{fig:p2_figure1}
\end{figure}

\paragraph{Framework Components}
\noindent
Using this foundation, we constructed \acrshort{framework} to include both an end-to-end phenotyping pipeline and a broader vision for next-generation computational phenotyping. To develop our end-to-end pipeline, we extended an existing framework originally developed for machine learning–based cohort discovery to more broadly capture computational phenotyping tasks.\cite{carrell_general_2024} Specifically, we added tasks for developing (\textit{Outline Phenotyping Heuristics}) and implementing (\textit{Apply Algorithm}) the phenotyping algorithm since they were previously implicit in development of the machine learning model. The end-to-end pipeline is detailed in Table \ref{tab:p2_table1}. Meanwhile, the broader overview indicates overall progress towards optimal human-machine synergy in next-generation phenotyping (Figure \ref{figure_1}). In this study, we only conducted a feasibility assessment of \acrshort{llms} for specific phenotyping tasks.

\subsubsection{Application of SHREC to Phenotyping Use Case}

\paragraph{Phenotyping Use Case}
\noindent
We leveraged computable phenotypes for \acrfull{arf} respiratory support therapies to demonstrate \acrshort{llm}-based methods for phenotyping tasks. Encounters were phenotyped based on the type and order of respiratory therapies received during individual \acrfull{icu} encounters.\cite{essay_rule-based_2020} The phenotypes were $1$) \acrfull{imv} only; $2$) \acrfull{nippv} only; $3$) \acrfull{hfni} only; $4$) \acrshort{nippv} Failure (or \acrshort{nippv} to \acrshort{imv}); $5$) \acrshort{hfni} Failure (or \acrshort{hfni} to \acrshort{imv}); $6$) \acrshort{imv} to \acrshort{nippv}; and $7$) \acrshort{imv} to \acrshort{hfni}.

\begin{table}[ht]
\vspace{10pt}
\caption{Overview of the individual end-to-end computational phenotyping tasks for \acrshort{framework} (\acrlong{framework}), a framework for next-generation computational phenotyping with \acrfull{llm}-based methods. Some tasks were based on a previously developed framework for phenotype development using machine learning algorithms.\cite{carrell_general_2024}}
\label{tab:p2_table1}
{\fontsize{10}{12}\selectfont
\renewcommand{\arraystretch}{1.25}
\begin{tabular}{>{\centering\arraybackslash}p{0.05\linewidth} | >{\raggedright\arraybackslash}p{0.20\linewidth} | >{\raggedright\arraybackslash}p{0.44\linewidth} | >{\raggedright\arraybackslash}p{0.21\linewidth}}
\hline
\textbf{Step} & \textbf{Name} & \textbf{Description} & \textbf{Analog to Previous Framework} \\
\hline

1 & Assess Fitness-for-Purpose & Determine the clinical outcome of interest, assess clinical significance, and assess any sources of clinical or data complexity. & Assess Fitness-for-Purpose \\
\hline

2 & Identify Data Sources & Identify data source(s) to use for phenotype development and evaluation. & Assess Fitness-for-Purpose \\
\hline

3 & Gather Ground Truth Data & Determine ground truth labels to use for validation of the phenotyping algorithm. & Create Gold Standard Data \\
\hline

4 & Outline Phenotyping Heuristics & Determine the tasks necessary in the phenotyping process to obtain the resulting phenotypes from the input data. Will likely include the inclusion and exclusion criteria. & None \\
\hline

5 & \acrshort{llm} Model Selection & If used, determine which \acrshort{llms} can be tested for specific tasks and how these models can be evaluated.\cite{pungitore_pheona_2025} For studies not using \acrshort{llms}, can either skip or identify other machine learning or \acrfull{nlp} models. & Develop Models \\
\hline

6 & Concept Selection & Classify or identify relevant data elements from the \acrfull{ehr} data. & Engineer Features \\
\hline

7 & Apply Algorithm & Apply the algorithm tasks to each record and identify the appropriate phenotype. & None \\
\hline

8 & Evaluate Algorithm & Determine the effectiveness of the phenotyping algorithm against the ground truth data. & Evaluate Models \\
\hline
\end{tabular}
}
\end{table}

\begin{figure}[ht]
  \centering
    \caption{Comparison of traditional and next-generation methods for constructing phenotypes for \acrfull{arf} respiratory support therapies, using the end-to-end phenotyping pipeline from \acrshort{framework} (\acrlong{framework}). The traditional methods were used for initial phenotype development\cite{essay_rule-based_2020} while the next-generation methods were implemented in this study.}    \includegraphics[width=0.85\linewidth]{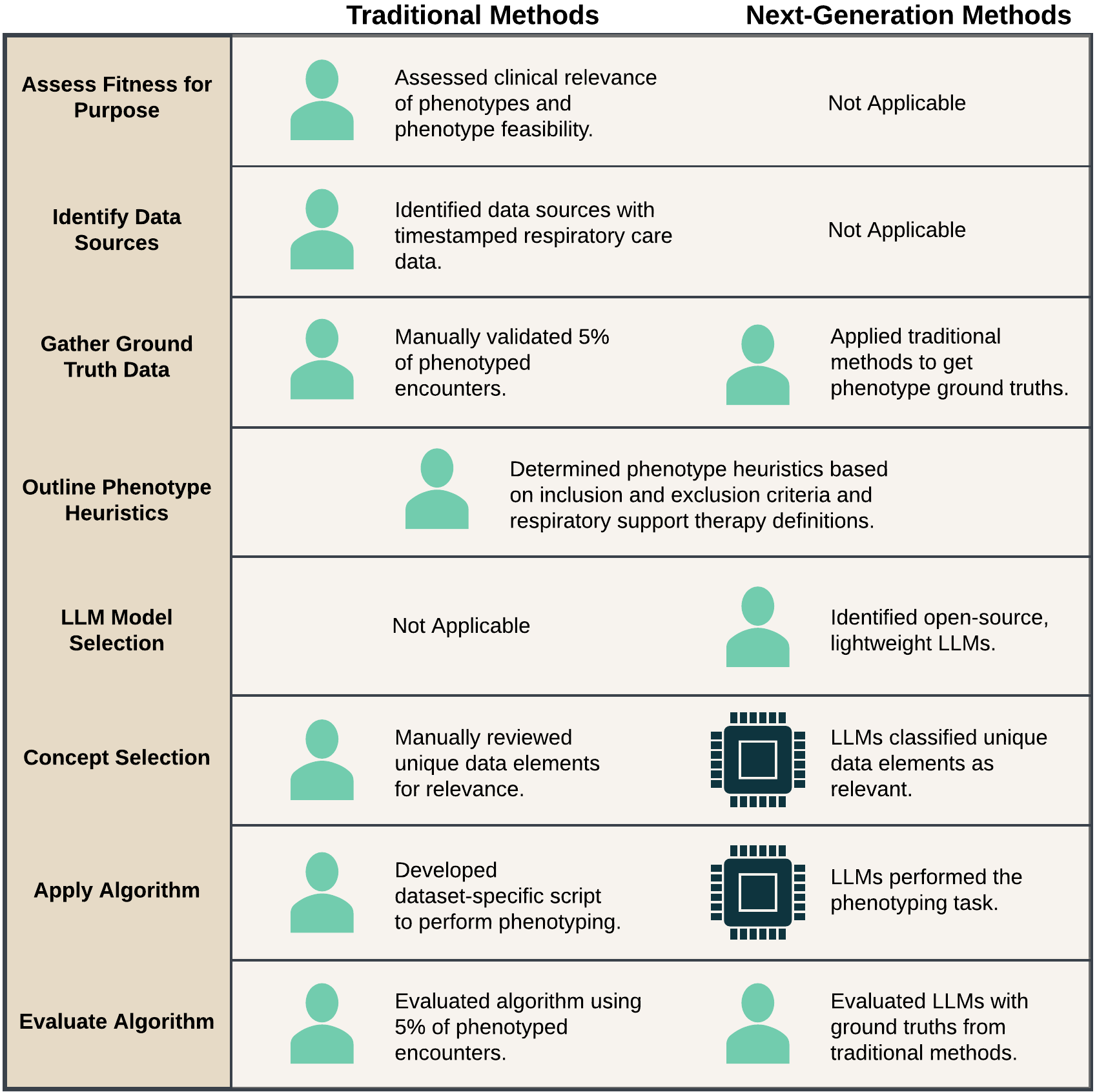}
    \label{fig:p2_figure2}
\end{figure}

\paragraph{Identification of Phenotyping Tasks}
\noindent
A comparison of the methods performed in the original study and this study are presented in Figure \ref{fig:p2_figure2}. In the original study, data from the \acrfull{eicu}\cite{pollard_eicu_2018} were manually reviewed to first determine relevance to the therapies or medications of interest and second to produce a phenotyping algorithm.\cite{essay_rule-based_2020} These processes mapped to the \textit{Concept Selection} and \textit{Apply Algorithm} tasks of the end-to-end phenotyping pipeline within \acrshort{framework}, respectively. In this study, we used \acrshort{llms} to perform the tasks of \textit{Concept Selection} and \textit{Apply Algorithm} while we also implemented \textit{Outline Phenotype Heuristics}, \textit{\acrshort{llm} Model Selection}, and \textit{Evaluate Algorithm} because they were required to execute and test the \acrshort{llm}-based methods. Since the remaining tasks were not related to the \acrshort{llm} methods, they were not included in this study, but were previously discussed in-depth.\cite{essay_rule-based_2020, carrell_general_2024}

\subsubsection{Implementation of LLMs for Phenotyping Tasks}
The following sections detail the methods for each of the implemented phenotyping tasks from \acrshort{framework}.

\paragraph{Outline Phenotype Heuristics}
\noindent
We determined the following heuristics from the previously developed algorithm:\cite{essay_rule-based_2020}

\begin{enumerate}
    \item Identified the first encounter for each unique patient. Removed additional encounters to ensure a single encounter per patient.
    
	\item Removed individuals less than $18$ years old at the start of the encounter.
	
	\item Extracted concepts from all distinct \acrshort{ehr} records across all encounters and determined which were relevant to the respiratory support therapies (\acrshort{imv}, \acrshort{nippv}, or \acrshort{hfni}) or medications (see Step $5a$) of interest. For example, ``\textit{BiPAP/CPAP}" indicates \acrshort{nippv} and ``\textit{Hi Flow NC}" indicates \acrshort{hfni}.

    \item Filtered records for each encounter to only the extracted concepts for all of the respiratory support therapies and medications of interest.
	
	\item Identified which of the respiratory support therapies were received during the encounter. The following criteria were used to determine if a therapy was received:
	
	\begin{enumerate}
		\item \acrshort{imv}: The presence of at least two records indicating \acrshort{imv} and at least one record indicating use of specific medication related to pre-intubation, intra-intubation, and post-intubation care (e.g., rapid sequence intubation medications, neuromuscular blocking agents, or continuous sedative agents).
		
		\item \acrshort{nippv}: At least two records indicating use of \acrshort{nippv} AND no records indicating use of \acrshort{hfni}.
		
		\item \acrshort{hfni}: The criteria for \acrshort{nippv} is met AND there is at least one additional record indicating use of \acrshort{hfni}.
	\end{enumerate}
	
	\item Determined the start and end for each treatment based on the offset time from \acrshort{icu} admission. When applicable, removed \acrshort{nippv} and \acrshort{hfni} records that occurred between \acrshort{imv} records from consideration and reassessed whether criteria for \acrshort{nippv} or \acrshort{hfni} was still met.
	
	\item Classified any encounters where any of the respiratory support therapies were received into one of the following $8$ phenotypes based on treatment criteria and ordering: $1$) \acrshort{imv} only; $2$) \acrshort{nippv} only; $3$) \acrshort{hfni} only; $4$) \acrshort{nippv} Failure (or \acrshort{nippv} to \acrshort{imv}); $5$) \acrshort{hfni} Failure (or \acrshort{hfni} to \acrshort{imv}); $6$) \acrshort{imv} to \acrshort{nippv}; $7$) \acrshort{imv} to \acrshort{hfni}; and $8$) No Therapies Received.
\end{enumerate}

\begin{table}[ht]
\vspace{10pt}
\caption{Constructed concept pattern for each selected table in the \acrfull{eicu} database.\cite{pollard_eicu_2018} Italicized text was replaced with values from the relevant column in each table, as demonstrated in the example for each table.}
\label{tab:p2_table2}
{
\renewcommand{\arraystretch}{1.05}
\begin{tabular}{>{\raggedright\arraybackslash}p{0.13\linewidth} | >{\raggedright\arraybackslash}p{0.41\linewidth} | >{\raggedright\arraybackslash}p{0.38\linewidth}} 
	
\hline
\textbf{Table Name} & \textbf{Constructed Concept Pattern} & \textbf{Example Constructed Concept} \\ 
\hline

Care Plan General & Source = Care Plan General; Concept = \textit{cplgroup}: \textit{cplitemvalue} & Source = Care Plan General; Concept = Route-Status: Oral - low sodium \\
\hline

Infusion Drug & Source = Infusion Drug; Concept = \textit{drugname} & Source = Infusion Drug; Concept = Amiodarone (mg/min) \\ 
\hline

Medication & Source = Medication; Concept = \textit{drugname} & Source = Medication; Concept = LOPRESSOR \\ 
\hline

Note & Source = Note; Concept = \textit{notevalue}: \textit{notetext} & Source = Note; Concept = denies fevers: denies fevers \\ 
\hline

Nurse Care & Source = Nurse Care; Concept = \textit{cellattributevalue} & Source = Nurse Care; Concept = emergency equipment at bedside \\ 
\hline

Nurse Charting & Source = Nurse Charting; Concept = \textit{nursingchartcelltypevalname}: \textit{nursingchartvalue} & Source = Nurse Charting; Concept = O2 Admin Device: BiPAP/CPAP \\ 
\hline

Respiratory Care & Source = Respiratory Care; Concept = \textit{airwaytype} & Source = Respiratory Care; Concept = Oral ETT \\ 
\hline

Respiratory Charting & Source = Respiratory Charting; Concept =  \textit{respcharttypecat}: \textit{respchartvaluelabel}: \textit{respchartvalue} & Source = Respiratory Charting; Concept = respFlowPtVentData: SaO2: 25 \\ 
\hline

Treatment & Source = Treatment; Concept = \textit{treatmentstring} & Source = Treatment; Concept = cardiovascular|myocardial ischemia / infarction|antiplatelet agent|aspirin \\ 

\hline

\end{tabular}
}
\end{table}

\paragraph{LLM Model Selection}
\noindent
To promote reproducibility and adaptability of our methods, we selected \acrshort{llms} available at the time of this study from Ollama, an open-source package that establishes local connections with open-source models.\cite{noauthor_ollamaollama_2024} Due to \acrfull{gpu} constraints, we selected the following lightweight, instruction-tuned models for testing: Mistral Small $24$ billion with Q$8.0$ quantization (model tag: 20ffe5db0161), Phi-$4$ $14$ billion with Q$8.0$ quantization (model tag: 310d366232f4), and Gemma$2$ $27$ billion with Q$8.0$ quantization (model tag: dab5dca674db).\cite{noauthor_ollamaollama_2024} DeepSeek-r1 $32$ billion with Q$4$\_K\_M quantization (model tag: 38056bbcbb2d) was previously tested for sampled data but was not used in this study due to high response latencies.\cite{noauthor_ollamaollama_2024, pungitore_pheona_2025} Models were run on a single Nvidia V$100$ $32$GB \acrshort{gpu}. Temperature and top-p were $0.0$ and $0.99$, respectively, for all experiments to avoid responses outside the requested format.

\paragraph{Concept Selection}
\noindent
For \textit{Concept Selection}, we generated constructed concepts from \acrshort{eicu} tables. Tables that did not include timestamped data but contained information on respiratory therapies or medications (such as the \textit{apacheApsVar} table) or were unlikely to contain descriptions of respiratory therapies or medications (such as the \textit{vitalPeriodic} table) were not processed further. There were $9$ tables used to construct input concepts (Table \ref{tab:p2_table2}). In early testing, we achieved best results when classifying the respiratory therapies separately from the relevant medications. We thus developed two prompts, resulting in two \acrshort{llm} responses per constructed concept (Supplementary Material). Concept definitions within the prompts were produced by two clinician experts who summarized and consolidated notes related to the terms of interest (Supplementary Material). We used \acrfull{cot} prompting by including a series of questions and answers to help the model determine relevancy of the constructed concept. \acrshort{cot} is a prompt engineering technique that has improved performance of \acrshort{llms} by using a series of reasoning tasks to guide the model to the final response.\cite{reynolds_prompt_2021, wei_chain--thought_2022} The answer to the final question for each prompt was parsed using string methods to get the final response of ``YES" or ``NO" for whether the constructed concept was relevant. Since there were two prompts, the final response was ``YES" if at least one of the responses was ``YES" and ``NO" otherwise.

\paragraph{Apply Algorithm}
\noindent
After we identified the relevant constructed concepts across the \acrshort{eicu} dataset, we applied the phenotyping heuristics. We identified the first encounter for each unique patient and then removed individuals who were less than $18$ years old at the start of the encounter. For each encounter, we filtered data from the original $9$ tables to the selected constructed concepts and then ordered each distinct constructed concept by its first occurrence based on the encounter admission time. The data, or constructed descriptions, for phenotyping were created by inserting each individual constructed concept into a string template and concatenating all of the unique constructed concepts together for the prompt. The template was \textit{``\#: \{constructed concept\}"} where ``\#" was the order of the constructed concept based on its first occurrence in the encounter records. Since timestamps were generalized to concept order and there was no encounter-specific information in the constructed descriptions, we phenotyped the unique constructed descriptions and then mapped the phenotypes to the relevant encounters for evaluation. Our phenotyping prompt used \acrshort{cot} (Supplementary Material). We parsed the answer to the final question to identify the selected phenotype.

\paragraph{Evaluate Algorithm}
\noindent
Models were evaluated for both \textit{Concept Selection} and \textit{Apply Algorithm} using components of \acrshort{pheona}, an evaluation framework for \acrshort{llm}-based applications to computational phenotyping.\cite{pungitore_pheona_2025} Previously, we evaluated the models for \textit{Concept Selection} using a random sample of constructed concepts.\cite{pungitore_pheona_2025} In this study, we evaluated these models on the full set of constructed concepts using \textit{Accuracy} (the ability of the model to produce accurate results) as the primary evaluation criterion and \textit{Model Response Latency} (how quickly model results were returned) as the secondary evaluation criterion from \acrshort{pheona}. We used \acrlong{roc} curve (\acrshort{roc}) to measure response accuracy against the concept ground truths. We measured response latency as the seconds required for the model to return a response for each constructed concept and then averaged these values for each prompt. For \textit{Apply Algorithm}, since we had not previously used \acrshort{pheona} for model evaluation, we evaluated model performance on a randomly selected subsample (Supplementary Material) and then evaluated the best performing models on all encounters. We also used \textit{Accuracy} and \textit{Model Response Latency} criteria to assess model performance with the \acrshort{roc} (and additionally, sensitivity and specificity) calculated using the original encounter ground truths.\cite{essay_rule-based_2020}

\subsection{Results}
\subsubsection{Phenotyping Use Case}
There were initially $200{,}859$ encounters across $166{,}355$ patients. After applying the inclusion and exclusion criteria, there were $159{,}701$ encounters for $159{,}701$ patients. Using the previously developed phenotyping algorithm, the encounters were phenotyped as follows: $16{,}736$ ($10.5\%$) as \acrshort{imv} only; $6{,}833$ ($4.3\%$) as \acrshort{nippv} only; $1{,}089$ ($0.7\%$) as \acrshort{hfni} only; $1{,}466$ ($0.9\%$) as \acrshort{nippv} Failure; $568$ ($0.4\%$) as \acrshort{hfni} Failure; $601$ ($0.4\%$) as \acrshort{imv} to \acrshort{nippv}; $186$ ($0.1\%$) as \acrshort{imv} to \acrshort{hfni}; and $132{,}222$ ($82.8\%$) as None.\cite{essay_rule-based_2020}

\subsubsection{Implementation of LLMs for Phenotyping Tasks}
\paragraph{Concept Selection}
\noindent
There were $572$ concept ground truths ($404$ \acrshort{arf} respiratory support therapies and $168$ medications) from the original phenotyping study.\cite{essay_rule-based_2020} Classification results based on the concept ground truths are presented in Table \ref{tab:p2_table3}. Mistral had the highest accuracy with an \acrshort{roc} of $0.896$ for classification of all concepts; however, it also had the highest total average latency ($26.2$ seconds compared to $22.0$ seconds and faster). All models performed better at medication classification (\acrshort{roc} of $0.997$ and higher) when compared to respiratory support therapy classification ($0.765$ and higher).

\begin{table}[ht]
\vspace{10pt}
\caption{Results of \textit{Concept Selection} using Gemma$2$ $27$ billion, Mistral Small $24$ billion, and Phi-$4$ $14$ billion \acrfull{llm} models. The number of concepts selected, \acrlong{roc} curve (\acrshort{roc}), and average latency (measured in seconds) were measured for both the \acrfull{arf} respiratory support therapies and medications prompts.}
\label{tab:p2_table3}
{\fontsize{10}{12}\selectfont
\renewcommand{\arraystretch}{1.25}
\begin{tabular}{
>{\centering\arraybackslash}p{0.09\linewidth} 
>{\centering\arraybackslash}p{0.05\linewidth} 
>{\centering\arraybackslash}p{0.10\linewidth} 
>{\centering\arraybackslash}p{0.07\linewidth} 
>{\centering\arraybackslash}p{0.05\linewidth} 
>{\centering\arraybackslash}p{0.10\linewidth} 
>{\centering\arraybackslash}p{0.07\linewidth} 
>{\centering\arraybackslash}p{0.05\linewidth} 
>{\centering\arraybackslash}p{0.09\linewidth} 
>{\centering\arraybackslash}p{0.07\linewidth}
}
\hline
\multirow{2}{*}{\textbf{Model}} &
\multicolumn{3}{c}{\textbf{Total Concepts}} &
\multicolumn{3}{c}{\makecell{\textbf{\acrshort{arf} Support Therapies} \\ \textbf{Concepts}}} &
\multicolumn{3}{c}{\makecell{\textbf{Medications} \\ \textbf{Concepts}}} \\
\cline{2-10}
& \textbf{N} & \textbf{AUROC$^a$} & \textbf{Latency} 
& \textbf{N} & \textbf{AUROC$^a$} & \textbf{Latency}
& \textbf{N} & \textbf{AUROC$^a$} & \textbf{Latency} \\
\hline

Gemma & $30{,}062$ & $0.792$ & $22.0$ & $29{,}394$ & $0.783$ & $17.7$ & $674$ & $0.997$ & $4.4$ \\
Mistral & $7{,}143$ & $0.896$ & $26.2$ & $6{,}754$ & $0.872$ & $16.6$ & $389$ & $0.996$ & $9.6$ \\
Phi & $13{,}829$ & $0.809$ & $19.0$ & $13{,}397$ & $0.765$ & $12.2$ & $433$ & $0.998$ & $6.8$ \\
\hline

\multicolumn{10}{l}{$^a$ \acrshort{roc}: Area under the receiver operating characteristic curve.}
\end{tabular}
}
\end{table}

\paragraph{Apply Algorithm}
\noindent
There were $97{,}583$ unique constructed descriptions for Gemma, $62{,}499$ for Mistral, and $65{,}581$ for Phi. However, since Gemma underperformed on the subsample of constructed descriptions (Supplementary Material), only Mistral and Phi were tested on the entire dataset. Mistral had an average response latency of $27.3$ seconds and Phi of $20.8$ seconds across all constructed descriptions. The \acrshort{roc}, sensitivity, and specificity for each phenotype are presented in Table \ref{tab:p2_table4}. Overall, Mistral performed better with a higher \acrshort{roc} on all phenotypes when compared to Phi. Mistral performed best on no therapy and single therapy phenotypes (None and \acrshort{imv}, \acrshort{nippv}, and \acrshort{hfni} Only) with an average \acrshort{roc} of $0.853$ while it only achieved an average \acrshort{roc} of $0.604$ on the remaining, multi-therapy phenotypes. Both models also had nearly perfect specificity for all phenotypes except \acrshort{imv} Only.

\subsection{Discussion}
In this study, we introduced \acrshort{framework}, a framework for applying \acrshort{llm}-based methods to computational phenotyping. We outlined the components of \acrshort{framework} and demonstrated how \acrshort{llms} can be used for computational phenotyping tasks.

\subsubsection{Development and Application of SHREC}
The primary contribution of this study was the components of \acrshort{framework}. For the first component, the end-to-end phenotyping pipeline, we expanded upon a previously developed framework and generalized it to apply to all computable phenotypes, not just those involving either machine learning or \acrshort{llms}.\cite{carrell_general_2024} For the second component, we outlined a novel system of \acrshort{llm}-based next-generation computational phenotyping. In its future state, we envision all repetitive tasks (including manual data review) being offloaded to \acrshort{llm} agents\cite{qiu_llm-based_2024} while humans guide complex tasks and provide deliberate oversight to best satisfy the \textit{Fundamental Theorem of Informatics}.\cite{friedman_fundamental_2009} Towards this end, we noted several key properties of \acrshort{llm}-based methods that would support widespread integration into the end-to-end pipeline. First, prompt engineering alone was sufficient for adapting the models to both \textit{Concept Selection} and \textit{Apply Algorithm} without additional retraining or algorithm development. Second, minimal data processing was required: other than tagging concepts with the original table name and recording order, raw \acrshort{ehr} data were used for both tasks. Therefore, even the lightweight models tested in this study demonstrated clear advantages over traditional phenotyping methods, including advanced \acrshort{nlp} and machine learning algorithms.

\subsubsection{Implementation of LLMs for Phenotyping Tasks}
The second contribution of this study was the demonstration of \acrshort{llms} for the tasks of \textit{Concept Selection} and \textit{Apply Algorithm} from the end-to-end pipeline. All models performed well at concept classification, especially classification of the medications concepts (Table \ref{tab:p2_table3}). For the phenotyping tasks, Mistral and Phi generally performed better at determining phenotypes with only a single treatment when compared to those with a sequence of treatments (Table \ref{tab:p2_table4}). We suspect the layered thought process of assigning records to a treatment and then determining treatment order was too complex for the models tested. We hypothesize that either mapping each constructed concept to a specific treatment or performing a second phenotyping step solely for treatment ordering would improve phenotyping performance. These results suggest that lightweight \acrshort{llms} can be readily applied to concept classification and simple phenotypes but may currently be insufficient for complex phenotypes without enhancements to the base models, prompts, or pipeline within \textit{Apply Algorithm}.

\begin{table}[ht]
\vspace{10pt}
\caption{Results of \textit{Apply Algorithm} using Mistral Small $24$ billion and Phi-$4$ $14$ billion  \acrfull{llm} models. The number of phenotyped encounters, \acrlong{roc} curve (\acrshort{roc}), sensitivity, and specificity were measured for both models across $159,701$ encounters.}
\label{tab:p2_table4}
{\fontsize{9}{11}\selectfont
\renewcommand{\arraystretch}{1.05}
\begin{tabular}{
>{\raggedright\arraybackslash}p{0.17\linewidth}  
>{\centering\arraybackslash}p{0.06\linewidth}    
>{\centering\arraybackslash}p{0.05\linewidth}    
>{\centering\arraybackslash}p{0.09\linewidth}    
>{\centering\arraybackslash}p{0.05\linewidth}    
>{\centering\arraybackslash}p{0.07\linewidth}    
>{\centering\arraybackslash}p{0.05\linewidth}    
>{\centering\arraybackslash}p{0.09\linewidth}    
>{\centering\arraybackslash}p{0.05\linewidth}    
>{\centering\arraybackslash}p{0.07\linewidth}    
}
\hline
\textbf{Phenotype} & \textbf{N} & \multicolumn{4}{c}{\textbf{Mistral}} & \multicolumn{4}{c}{\textbf{Phi}} \\
\cline{3-10}
& & \textbf{N} & \textbf{AUROC$^a$} & \textbf{Sens.$^b$} & \textbf{Spec.$^c$} & \textbf{N} & \textbf{AUROC$^a$} & \textbf{Sens.$^b$} & \textbf{Spec.$^c$} \\
\hline
\acrshort{imv}$^d$ Only & $16{,}736$ & $33{,}537$ & $0.881$ & $0.898$ & $0.864$ & $36{,}343$ & $0.863$ & $0.936$ & $0.791$ \\

\acrshort{nippv}$^e$ Only & $6{,}833$ & $7{,}060$ & $0.809$ & $0.637$ & $0.981$ & $7{,}158$ & $0.758$ & $0.547$ & $0.969$ \\

\acrshort{hfni}$^f$ Only & $1{,}089$ & $2{,}325$ & $0.825$ & $0.661$ & $0.989$ & $843$ & $0.543$ & $0.093$ & $0.994$ \\

\acrshort{nippv}$^e$ Failure & $1{,}466$ & $2{,}063$ & $0.717$ & $0.443$ & $0.991$ & $1{,}463$ & $0.669$ & $0.347$ & $0.992$ \\

\acrshort{hfni}$^f$ Failure & $568$ & $377$ & $0.513$ & $0.028$ & $0.998$ & $208$ & $0.503$ & $0.007$ & $0.998$ \\

\acrshort{imv}$^d$ to \acrshort{nippv}$^e$ & $601$ & $1{,}725$ & $0.526$ & $0.063$ & $0.989$ & $1{,}536$ & $0.565$ & $0.143$ & $0.987$ \\

\acrshort{imv}$^d$ to \acrshort{hfni}$^f$ & $186$ & $1{,}545$ & $0.659$ & $0.328$ & $0.990$ & $1{,}026$ & $0.539$ & $0.086$ & $0.991$ \\

None & $132{,}222$ & $103{,}990$ & $0.896$ & $0.824$ & $0.968$ & $66{,}952$ & $0.845$ & $0.744$ & $0.947$ \\
\hline
\multicolumn{10}{l}{$^a$ \acrshort{roc}: Area under the receiver operating characteristic curve.} \\
\multicolumn{10}{l}{$^b$ Sens.: Sensitivity} \\
\multicolumn{10}{l}{$^c$ Spec.: Specificity} \\
\multicolumn{10}{l}{$^d$ \acrshort{imv}: \acrlong{imv}.} \\
\multicolumn{10}{l}{$^e$ \acrshort{nippv}: \acrlong{nippv}.} \\
\multicolumn{10}{l}{$^f$ \acrshort{hfni}: \acrlong{hfni}.}
\end{tabular}
}
\end{table}

One outstanding question for all biomedical tasks performed with \acrshort{llms} is how to best incorporate specialized medical knowledge, including standardized vocabularies and ontologies. In this study, we injected medical information into the prompts and relied on the inherent capabilities of each model for data synthesis and comprehension. This method is applicable to many other computable phenotypes, including the previously discussed phenotypes for \acrshort{pasc} where the \acrfull{omop} concepts could be categorized by injecting prompts with symptom information.\cite{pungitore_computable_2024} Outside of computational phenotyping, studies have also explored methods for providing \acrshort{snomed} knowledge to \acrshort{llms}, including prompt injection, model pretraining, and model finetuning, although almost none of these studies reported performance results after incorporating \acrshort{snomed}.\cite{chang_use_2024} Furthermore, a recent study found finetuned biomedical \acrshort{llms} underperformed when compared to generalist models on multiple clinical benchmarking tasks.\cite{dorfner_evaluating_2025} Therefore, while there is some evidence that prompt injection (with or without \acrlong{rag}) may be the best method for incorporating domain knowledge, our results suggest there may be limits to the effectiveness of this method. Thus, there remain significant gaps in understanding how to best incorporate specialized medical knowledge into \acrshort{llms} both for general biomedical and computational phenotyping tasks.

\subsubsection{Study Limitations}
There were several limitations to the framework and methods implemented in this study. First, repetitive manual review was still required for model evaluation. Furthermore, since we used previously developed ground truths, we performed less manual review than would be required for a novel phenotyping study. However, manual review for \acrshort{llm}-based methods can be reduced by reviewing samples rather than the entire dataset. Another limitation is scalability. For \textit{Concept Selection} and \textit{Apply Algorithm}, each constructed concept and description required individual \acrshort{llm} responses and thus, after a certain number of records, it would become infeasible to use \acrshort{llms}. However, for this study, almost a sixfold increase in records would be required before the time for \acrshort{llm}-based phenotyping became comparable to the time for traditional phenotyping methods. Additionally, as improvements in model performance, \acrlong{rag} architecture, and prompt engineering are developed, we expect the issue of scalability for \acrshort{llm}-based methods to be lessened, although not completely mitigated.

\subsubsection{Future Directions}
There are many future directions to explore outside of those highlighted by the broader vision of computational phenotyping from \acrshort{framework}. One direction is to understand how \acrshort{llms} reason with respect to computational phenotyping. Although we used \acrshort{cot} because of its demonstrated ability to produce accurate results,\cite{reynolds_prompt_2021, wei_chain--thought_2022, zhang_igniting_2025} recent studies have suggested that \acrshort{cot} reasoning may actually be unfaithful.\cite{chen_reasoning_2025, arcuschin_chain--thought_2025, turpin_language_2023} Given the complexity of phenotyping tasks, we suggest studying \acrshort{cot} reasoning to understand when and how logical inconsistencies may arise. Another future direction is to adapt phenotype definitions to \acrshort{llms}. For example, the clinician experts indicated some of the medications may be present for noninvasive therapies, but not as a continuous infusion (Supplementary Material). However, for the previously developed definition, only medication presence was used due to complexities in algorithmically processing medication information in relation to respiratory therapies. In future iterations, we would update the phenotype definition to include administration method rather than simply asking the model to look for concept presence. Finally, we propose development of industry standards for evaluation of \acrshort{llm}-based methods specifically with respect to automated processes to ensure appropriate oversight.

\subsection{Conclusion}
We developed \acrshort{framework}, a framework that describes how to apply \acrshort{llm}-based methods to computational phenotyping. \acrshort{framework} outlines both an end-to-end pipeline for computable phenotype development along with a broader vision of next-generation phenotyping using \acrshort{llm}-based methods. We demonstrated \acrshort{framework} on a phenotyping use case to assess the feasibility of \acrshort{llms} for specific phenotyping tasks and promote further research into next-generation computational phenotyping methods. This work is applicable to all computational phenotyping studies, particularly those using manual review for phenotype development.

\vspace{5pt}
\addcontentsline{toc}{subsection}{References}
\begin{refcontext}[sorting=none]
\printbibliography[heading=subbibliography]
\end{refcontext}

\end{refsection}

\end{document}